\documentclass[12pt]{article}
\usepackage{amsmath}
\usepackage{amssymb}

\DeclareMathOperator{\csch}{csch}
\DeclareMathOperator{\sech}{sech}

\title{
Exceptional orthogonal polynomials and new exactly solvable potentials in quantum mechanics}
\author{C. QUESNE\\
{\small \sl Physique Nucl\'eaire Th\'eorique et Physique Math\'ematique,}\\ 
{\small \sl Universit\'e Libre de Bruxelles, Campus de la Plaine CP229,} \\ 
{\small \sl Boulevard~du Triomphe, B-1050 Brussels, Belgium} \\
{\small \sl cquesne@ulb.ac.be}}
\date{ }
\begin{document}
\baselineskip=22pt plus 1pt minus 1pt
\maketitle

\begin{abstract}
In recent years, one of the most interesting developments in quantum mechanics has been the construction of new exactly solvable potentials connected with the appearance of families of exceptional orthogonal polynomials (EOP) in mathematical physics. In contrast with families of (Jacobi, Laguerre and Hermite) classical orthogonal polynomials, which start with a constant, the EOP families begin with some polynomial of degree greater than or equal to one, but still form complete, orthogonal sets with respect to some positive-definite measure. We show how they may appear in the bound-state wavefunctions of some rational extensions of well-known exactly solvable quantum potentials. Such rational extensions are most easily constructed in the framework of supersymmetric quantum mechanics (SUSYQM), where they give rise to a new class of translationally shape invariant potentials. We review the most recent results in this field, which use higher-order SUSYQM. We also comment on some recent re-examinations of the shape invariance condition, which are independent of the EOP construction problem.
\end{abstract}

%
%
\section{Introduction}

As well known, the classical orthogonal polynomials (COP) of Jacobi, Laguerre and Hermite \cite{erdelyi} have a lot of applications in applied mathematics and in physics. Since the early days of quantum mechanics, in particular, they have made their appearance in the bound-state wavefunctions of exactly solvable (ES) potentials.\par
%
%
In this respect, the factorization method, initiated by Schr\"odinger \cite{schrodinger} and developed by Infeld and Hull \cite{infeld}, has played an important role. This technique has been reformulated in the framework of supersymmetric quantum mechanics (SUSYQM) \cite{cooper00}, also connected with the Darboux algorithm \cite{darboux}.\par
%
%
The concept of shape invariance (SI), introduced by Gendenshtein \cite{gendenshtein}, has enriched the set of tools available, as its combination with SUSYQM has provided an integration method for the Schr\"odinger equation. In such a context, the class of translationally SI potentials is especially interesting because it includes all well-known ES potentials found in most textbooks on quantum mechanics. With the addition of some potentials depending upon more than two parameters \cite{carinena00}, the classification given in \cite{cooper87, dabrowska} was until recently thought to be complete \cite{cooper00}.\par
%
%
In 2008, the introduction of the first families of exceptional orthogonal polynomials (EOP) \cite{gomez10a, gomez09} and the realization of their usefulness in constructing new ES quantum potentials \cite{cq08}, which were so far unknown translationally SI ones, produced a revival in this field.\par
%
%
The purpose of the present contribution is to review the fast-growing branch of EOP and related ES potentials.\par
%
%
\section{New families of orthogonal polynomials and quantum mechanics}

The COP are often characterized as the polynomial solutions of a Sturm-Liouville problem in connection with the celebrated theorem of Bochner \cite{bochner}. According to the latter, if an infinite sequence of polynomials $P_n(z)$, $n=0$, 1, 2,~\ldots, satisfies a second-order eigenvalue equation
\begin{equation}
  p(z) P_n''(z) + q(z) P_n'(z) + r(z) P_n(z) = \lambda_n P_n(z), \qquad n=0, 1, 2, \ldots, 
\end{equation}
then $p(z)$, $q(z)$ and $r(z)$ must be polynomials of degree 2, 1 and 0, respectively. In addition, if the sequence is an orthogonal polynomial system, then it has to be (up to an affine transformation) one of the COP of Jacobi, Laguerre or Hermite.\par
%
%
Many generalizations and extensions of these EOP families have been proposed, among which we may quote the Askey scheme of hypergeometric orthogonal polynomials and its $q$-analogue (see, e.g., \cite{koekoek}), as well as algebraic deformations of SI potentials \cite{gomez04a, gomez04b}.\par
%
%
More recently, the concept of an exceptional polynomial subspace and the closely related notion of EOP were introduced \cite{gomez10a}. As the COP, the latter are eigenfunctions of a second-order differential operator, but in contrast with Bochner's assumption, the first eigenpolynomial of the sequence is of degree $m \ge 1$, though the system of polynomials $P_n(z)$, $n=m$, $m+1$, $m+2$,~\ldots, denoted by $X_m$, still forms an orthogonal and complete set with respect to some positive-definite measure. The $X_1$ case was studied in detail in \cite{gomez10a, gomez09}, where it was shown that there exist two such systems of Jacobi or Laguerre type.\par
%
%
As an illustration, let us quote some properties of the latter \cite{gomez09}, denoted by $\hat{L}^{(\alpha)}_n(z)$, $n=1$, 2, 3,~\ldots,
\begin{equation}
  \hat{L}^{(\alpha)}_1(z) = - z - \alpha - 1, \qquad \hat{L}^{(\alpha)}_2(z) = z^2 - \alpha(\alpha+2), \ldots, 
  \label{eq:ex-X1} 
\end{equation}
\begin{equation}
  \left\{z \frac{d^2}{dz^2} - \frac{z-\alpha}{z+\alpha} \left[(z+\alpha+1) \frac{d}{dz} - 1\right]\right\}
  \hat{L}^{(\alpha)}_n(z) = - (n-1) \hat{L}^{(\alpha)}_n(z),  \label{eq:eq-X1}
\end{equation}
\begin{equation}
  \int_0^{\infty} \hat{L}^{(\alpha)}_{n'}(z) \hat{L}^{(\alpha)}_n(z) \frac{z^{\alpha} e^{-z}}{(z+\alpha)^2} dz
  = \delta_{n',n} \frac{\Gamma(n+\alpha+1)}{(n+\alpha-1) (n-1)!},  \label{eq:int-X1}
\end{equation}
and contrast them with corresponding properties \cite{erdelyi} of conventional Laguerre polynomials $L^{(\alpha)}_n(z)$, $n=0$, 1, 2,~\ldots,
\begin{equation}
  L^{(\alpha)}_0(z) = 1, \qquad L^{(\alpha)}_1(z) = - z + \alpha + 1, \ldots, 
\end{equation}
\begin{equation}
  \left[z \frac{d^2}{dz^2} + (\alpha+1-z) \frac{d}{dz}\right] L^{(\alpha)}_n(z) = - n L^{(\alpha)}_n(z),
\end{equation}
\begin{equation}
  \int_0^{\infty} L^{(\alpha)}_{n'}(z) L^{(\alpha)}_n(z) z^{\alpha} e^{-z} dz = \delta_{n',n} 
  \frac{\Gamma(n+\alpha+1)}{n!}.
\end{equation}
The former can actually be written as linear combinations of three of the latter,
\begin{equation}
  \hat{L}^{(\alpha)}_n(z) = n L^{(\alpha)}_n(z) - 2(n+\alpha) L^{(\alpha)}_{n-1}(z) + (n+\alpha) 
  L^{(\alpha)}_{n-2}(z).
\end{equation}
\par
%
%
Both $X_1$ families of EOP were then shown to have interesting applications in quantum mechanics \cite{cq08}. For such a purpose, use was made of the point canonical transformation (PCT) method \cite{bhatta, levai}, which consists in mapping a Schr\"odinger equation containing some physically acceptable potential into a second-order differential equation satisfied by some polynomials by making changes of variable and of function.\par
%
%
In the case of equation (\ref{eq:eq-X1}) for Laguerre-type EOP $\hat{L}^{(\alpha)}_n(z)$, the PCT method led to the Schr\"odinger equation for a rationally-extended radial oscillator (in units $\hbar = 2m = 1$)
\begin{equation}
  \left[- \frac{d^2}{dx^2} + V_{l,{\rm ext}}(x)\right] \hat{\psi}^{(l)}_{\nu}(x) = E^{(l)}_{\nu} \hat{\psi}^{(l)}_{\nu}(x), 
  \qquad 0 < x < \infty,
\end{equation}
where
\begin{equation}
  V_{l,{\rm ext}}(x) = V_l(x) + V_{l,{\rm rat}}(x), \qquad V_l(x) = \frac{1}{4} \omega^2 x^2 + \frac{l(l+1}{x^2},
  \label{eq:ext-osc}
\end{equation}
\begin{equation}
  V_{l,{\rm rat}}(x) = \frac{4\omega}{\omega x^2+2l+1} - \frac{8\omega(2l+1)}{(\omega x^2+2l+1)^2}.
  \label{eq:rat-osc}
\end{equation}
Here $V_l(x)$ is a conventional radial oscillator potential, where $\omega$ and $l$ denote the oscillator frequency and the angular momentum quantum number, respectively. The additional term $V_{l, {\rm rat}}(x)$ does not modify the behaviour of $V_l(x)$ for large values of $x$, while for small values it has only an effect for $l=0$ as it changes $V_0(0) = 0$ into $V_{0, {\rm ext}}(0) = - 4\omega$. No singularity being introduced on the half-line, the resulting potential $V_{l, {\rm ext}}(x)$ is a well-behaved quantum one.\par%
%
It turns out that $V_{l, {\rm ext}}(x)$ has the same bound-state spectrum as $V_l(x)$, namely
\begin{equation}
  E^{(l)}_{\nu} = \omega \bigl(2\nu + l + \tfrac{3}{2}\bigr), \qquad \nu=0, 1, 2, \ldots.
\end{equation}
Whereas the bound-state wavefunctions of the latter,
\begin{equation}
  \psi^{(l)}_{\nu}(x) \propto x^{l+1} e^{- \frac{1}{4} \omega x^2} L^{(l + \frac{1}{2})}_{\nu}(\tfrac{1}{2} \omega
  x^2) \propto \eta_l(z) L^{(\alpha)}_{\nu}(z), \qquad \nu = 0, 1, 2, \ldots,  \label{eq:osc-wf}
\end{equation}
\begin{equation}
  z = \tfrac{1}{2} \omega x^2, \qquad \alpha = l + \tfrac{1}{2}, \qquad \eta_l(z) = z^{\frac{1}{4}(2\alpha+1)}
  e^{- \frac{1}{2}z},  \label{eq:z}  
\end{equation}
can be expressed in terms of Laguerre polynomials, those of the former,
\begin{equation}
  \hat{\psi}^{(l)}_{\nu}(x) \propto \frac{x^{l+1} e^{- \frac{1}{4} \omega x^2}}{\omega x^2 + 2l+1} 
  \hat{L}^{(l + \frac{1}{2})}_{\nu+1}(\tfrac{1}{2} \omega x^2) \propto \frac{\eta_l(z)}{z+\alpha} 
  \hat{L}^{(\alpha)}_{\nu+1}(z), \qquad \nu = 0, 1, 2, \ldots,
\end{equation}
are given in terms of Laguerre-type $X_1$ EOP.\par
%
%
The same approach used for Jacobi-type $X_1$ EOP led to a rational extension $V_{A,B,{\rm ext}}(x)$ of the well-known Scarf I potential $V_{A,B}(x) = [A(A-1) + B^2] \sec^2 x - B (2A-1) \sec x \tan x$, $- \frac{\pi}{2} < x < \frac{\pi}{2}$, $0 < B < A-1$,\footnote{Note that some people prefer to use the so-called trigonometric P\"oschl-Teller (or P\"oschl-Teller I) potential $V_{\bar{A}, \bar{B}}(\bar{x}) = \bar{A}(\bar{A}-1) \sec^2 \bar{x} + \bar{B}(\bar{B}-1) \csc^2 \bar{x}$, $0 < \bar{x} < \frac{\pi}{2}$, related to the Scarf I by the changes of parameters and of variable $\bar{A} = A-B$, $\bar{B} = A+B$, $\bar{x} = \frac{1}{2}\left(x + \frac{\pi}{2}\right)$. By the same transformation, $V_{A,B,{\rm ext}}(x)$ becomes an extended trigonometric P\"oschl-Teller potential $V_{\bar{A}, \bar{B}, {\rm ext}}(\bar{x})$.} with similar properties \cite{cq08}.\par
%
%
{}Furthermore, it was proved \cite{cq08} that the two new ES potentials $V_{l, {\rm ext}}(x)$ and $V_{A, B, {\rm ext}}(x)$ were translationally SI or, in other words, that their SUSY partner in a first-order SUSYQM approach was a potential of the same type but with different parameters, namely $V_{l+1, {\rm ext}}(x)$ and $V_{A+1, B, {\rm ext}}(x)$, respectively (see section 5 for more details on SI).\par
%
%
The discovery of this surprising property showed that SUSYQM techniques might prove convenient to build new examples of ES potentials related to $X_1$ EOP, in the same spirit as previous works on algebraic deformations of SI potentials \cite{gomez04a, gomez04b}. This was confirmed by the construction of a rational extension $V_{A, B, {\rm ext}}(x)$ of the generalized P\"oschl-Teller potential $V_{A, B}(x) = [B^2 + A(A+1)] \csch^2 x  - B (2A+1) \csch x \coth x$, $0 < x < \infty$, $B > A+1 > 1$ \cite{bagchi09a}.\footnote{Note that the hyperbolic P\"oschl-Teller (or P\"oschl-Teller II) potential $V_{\bar{A}, \bar{B}}(\bar{x}) = - \bar{A} (\bar{A}+1) \sech^2 \bar{x} + \bar{B} (\bar{B}-1) \csch^2 \bar{x}$, $0 < \bar{x} < \infty$, is related to the generalized P\"oschl-Teller one by the changes of parameters and of variable $\bar{A} = A+B$, $\bar{B} = B-A$, $\bar{x} = \frac{1}{2} x$. As a consequence, $V_{A, B, {\rm ext}}(x)$ is connected with an extended hyperbolic P\"oschl-Teller potential $V_{\bar{A}, \bar{B}, {\rm ext}}(\bar{x})$, which can be derived by the same transformation.} In the same work, it was also pointed out that the extension procedure was not successful for any well-known ES potential owing to the possible appearance of a pole in the rational part. This was explicitly shown on the example of the Scarf II potential $V_{A, B}(x) = [B^2 - A(A+1)] \sech^2 x  + B (2A+1) \sech x \tanh x$, $- \infty < x < \infty$, $A > 0$. It was then suggested that $\cal PT$ symmetry might facilitate reconciling the construction of ES rational potentials to the requirement that these be singularity free.\par
%
%
With a similar approach, the first examples of Laguerre- and Jacobi-type $X_2$ EOP were built in connection with some new rationally-extended radial oscillator and Scarf I potentials \cite{cq09}. A striking feature was that in each case there appeared two distinct $X_2$ EOP in contrast with the case of $X_1$ where only one did exist.\par
%
%
The question of arbitrary large $m$ was successfully addressed by the construction of two distinct families of Laguerre- and Jacobi-type $X_m$ EOP \cite{odake09a, odake10a}, now labelled as type I and type II, and the thorough study of their properties \cite{odake10b, ho09}. It then became clear that in the $m=1$ special case the polynomials of the two families happened to be proportional, hence leaving only a single independent one.\par
%
%
Later on, the $X_m$ EOP were shown to be obtainable through several equivalent approaches to the SUSYQM procedure, such as the Darboux-Crum transformation \cite{gomez10b, gomez11a, sasaki}, the Darboux-B\"acklund one \cite{grandati11a} and the prepotential method \cite{ho11a}.\par
%
%
Very recently, the $X_m$ EOP (and related potentials) were generalized to multi-indexed families $X_{m_1, m_2, \ldots, m_k}$ by making use of multi-step Darboux algebraic transformations \cite{gomez11b}, the Crum-Adler mechanism \cite{odake11a}, higher-order SUSYQM \cite{cq11a, cq11b} or multi-step Darboux-B\"acklund transformations \cite{grandati11b}.\par
%
%
\section{\boldmath Rationally-extended radial oscillator and Laguerre-type $X_m$ EOP in first-order SUSYQM}

As a example, let us deal with the case of the radial oscillator potential $V_l(x)$, defined in (\ref{eq:ext-osc}), in the framework of first-order SUSYQM.\par
%
%
In such an approach, one considers a pair of SUSY partners \cite{cooper00}
\begin{equation}
\begin{split}
  & H^{(+)} = A^{\dagger} A = - \frac{d^2}{dx^2} + V^{(+)}(x) - E, \qquad H^{(-)} = A A^{\dagger} = 
       - \frac{d^2}{dx^2} + V^{(-)}(x) - E, \\
  & A^{\dagger} = - \frac{d}{dx} + W(x), \qquad A = \frac{d}{dx} + W(x), \qquad V^{(\pm)}(x) = W^2(x) \mp 
       W'(x) + E,  
\end{split}  \label{eq:SUSYQM}
\end{equation}
which intertwine with the first-order differential operators $A$ and $A^{\dagger}$ as $A H^{(+)} = H^{(-)} A$ and $A^{\dagger} H^{(-)} = H^{(+)} A^{\dagger}$. Here $W(x)$ is the superpotential, which can be expressed as $W(x) = - \bigl(\log \phi(x)\bigr)'$ in terms of a factorization function $\phi(x)$, $E$ is the factorization energy and a prime denotes a derivative with respect to $x$.\par%
%
The factorization function is a (nodeless) seed solution of the initial Schr\"odinger equation
\begin{equation}
  \left(- \frac{d^2}{dx^2} + V^{(+)}(x)\right) \phi(x) = E \phi(x)  \label{eq:phi-E}
\end{equation}
with energy $E$ smaller than or equal to the ground-state energy $E^{(+)}_0$ of $V^{(+)}$. One may distinguish three cases. In case i, $E = E^{(+)}_0$, $\phi(x) = \psi^{(+)}_0(x)$ and $V^{(-)}$ has the same spectrum as $V^{(+)}$ except for $E^{(+)}_0$, which is removed. In case ii, $E < E^{(+)}_0$, $\phi(x)$ and $\phi^{-1}(x)$ are both nonnormalizable and $V^{(-)}$ has exactly the same spectrum as $V^{(+)}$ (isospectral case). Finally, in case iii, $E < E^{(+)}_0$ and $\phi^{-1}(x)$ is normalizable (while $\phi(x)$ is of course nonnormalizable), so that $V^{(-)}$ has an extra bound state at an energy $E$ below the spectrum of $V^{(+)}$.\par
%
%
{}Furthermore, from the intertwining relations, it results that the wavefunctions of $V^{(-)}$ can be obtained from those of $V^{(+)}$ with the same energy by applying the operator $A$.\par
%
%
We shall be interested here in case ii (isospectral case). It turns out that for the radial oscillator potential $V_l(x)$, there are two types of (nodeless) seed functions, whose energy $E$ is less than $E^{(+)}_0 = \omega (l + \frac{3}{2})$ and whose inverse is nonnormalizable, namely
\begin{equation}
  \phi^{\rm I}_{lm}(x) = \chi^{\rm I}_l(z) L^{(\alpha)}_m(-z) \propto x^{l+1} e^{\frac{1}{4} \omega x^2} 
  L^{(l+\frac{1}{2})}_m(- \tfrac{1}{2} \omega x^2),  \label{eq:phi-I}
\end{equation}
\begin{equation}
  \phi^{\rm II}_{lm}(x) = \chi^{\rm II}_l(z) L^{(-\alpha)}_m(z) \propto x^{-l} e^{-\frac{1}{4} \omega x^2} 
  L^{(-l-\frac{1}{2})}_m(\tfrac{1}{2} \omega x^2),  \label{eq:phi-II}
\end{equation}
with
\begin{equation}
  \chi^{\rm I}_l(z) = z^{\frac{1}{4}(2\alpha+1)} e^{\frac{1}{2}z}, \qquad \chi^{\rm II}_l(z) = 
  z^{-\frac{1}{4}(2\alpha-1)} e^{-\frac{1}{2}z}, 
\end{equation}
and $z$, $\alpha$ defined in equation (\ref{eq:z}). The corresponding energies are $E^{\rm I}_{lm} = - \omega(\alpha + 2m + 1)$, $m=1$, 2, 3,~\ldots, and $E^{\rm II}_{lm} = - \omega(\alpha - 2m - 1)$, $m=1$, 2, \ldots, $l$ ($< \alpha$),  respectively.\par
%
%
To get as partner potential a rationally-extended potential with a given $l$, $V^{(-)}(x) = V_{l,{\rm ext}}(x) + C$, where $C$ is some additive constant, we have to start from a conventional radial oscillator potential $V_{l'}(x)$ with a different $l'$. The rational part of $V_{l,{\rm ext}}(x)$ (see equation (\ref{eq:ext-osc})) can then be expressed in terms of some $m$th-degree polynomial in $z$, $g^{(\alpha)}_m(z)$, as
\begin{equation}
  V_{l,{\rm rat}}(x) = - 2\omega \left\{\frac{\dot{g}^{(\alpha)}_m}{g^{(\alpha)}_m} + 2z 
  \left[\frac{\ddot{g}^{(\alpha)}_m}{g^{(\alpha)}_m} - \left(\frac{\dot{g}^{\alpha)}_m}
  {g^{(\alpha)}_m}\right)^2\right]\right\},  \label{eq:V-rat}
\end{equation}
where a dot denotes a derivative with respect to $z$. According to the choice made for $\phi(x)$, we may distinguish the two cases
\begin{eqnarray}
  & (i) \; &  l' = l-1, \quad \phi = \phi^{\rm I}_{l-1,m}, \quad g^{(\alpha)}_m(z) = L^{(\alpha-1)}_m(-z), \quad
         C = - \omega, \nonumber \\
  & & m=1, 2, 3, \ldots; \\
  & (ii) \; &  l' = l+1, \quad \phi = \phi^{\rm II}_{l+1,m}, \quad g^{(\alpha)}_m(z) = L^{(-\alpha-1)}_m(z), \quad
         C = \omega, \nonumber \\
  & & m=1, 2, \ldots, l+1 (< \alpha+1). 
\end{eqnarray}
From the properties of Laguerre polynomials \cite{erdelyi}, it follows that $V_{l,{\rm ext}}(x)$ has no pole on $(0, \infty)$.\par
%
%
{}For $m=1$, for instance, we get $g^{(\alpha)}_1(z) = z + \alpha$ for type I and $g^{(\alpha)}_1(z) = - z - \alpha$ for type II. Hence there is a single rationally-extended potential, as obtained in \cite{cq08, cq09} (see equations (\ref{eq:ext-osc}) and (\ref{eq:rat-osc})). In contrast, for $m=2$, $g^{(\alpha)}_2(z) = \frac{1}{2}[z^2 + 2 (\alpha+1) z + \alpha (\alpha+1)]$ or $g^{(\alpha)}_2(z) = \frac{1}{2}[z^2 + 2 (\alpha-1) z + \alpha (\alpha-1)]$ according to the choice made. Hence there are two distinct extended potentials, as first observed in \cite{cq09} and confirmed for all $m \ge 2$ in \cite{odake09a, odake10a}.\par
%
%
{}From the wavefunctions $\psi^{(+)}_{\nu}(x) \propto \eta_{l'}(z) L^{(\alpha')}_{\nu}(z)$, $\alpha' = l' + \frac{1}{2}$, $\nu=0$, 1, 2,~\ldots, of $V^{(+)}$, those of $V^{(-)}$ with the same energy are obtained as
\begin{equation}
  \psi^{(-)}_{\nu}(x) \propto A \psi^{(+)}_{\nu}(x) \propto \frac{\eta_l(z)}{g^{(\alpha)}_m(z)} 
  y^{(\alpha)}_n(z), \qquad n = m + \nu, \qquad \nu=0, 1, 2, \ldots,  \label{eq:partner-wf}
\end{equation}
where $y^{(\alpha)}_n(z)$ is some $n$th-degree polynomial in $z$. The latter results from the application of a first-order differential operator ${\cal O}^{(\alpha)}_m$ on the conventional Laguerre polynomial $L^{(\alpha')}_{\nu}(z)$,   
\begin{equation}
  y^{(\alpha)}_n(z) \propto {\cal O}^{(\alpha)}_m L^{(\alpha')}_{\nu}(z). 
\end{equation}
Here ${\cal O}^{(\alpha)}_m = g^{(\alpha)}_m \left(\frac{d}{dz} - 1\right) - \dot{g}^{(\alpha)}_m$ for type I and ${\cal O}^{(\alpha)}_m = g^{(\alpha)}_m \left(z\frac{d}{dz} + \alpha + 1\right) - z \dot{g}^{(\alpha)}_m$ for type II.\par
%
%
On the other hand, by directly inserting equation (\ref{eq:partner-wf}) in the Schr\"odinger equation for $V^{(-)}(x)$, one arrives at the following second-order differential equation satisfied by $y^{(\alpha)}_n(z)$,
\begin{equation}
  \left[z \frac{d^2}{dz^2} + \left(\alpha + 1 - z - 2z \frac{\dot{g}^{(\alpha)}_m}{g^{(\alpha)}_m}\right)
  \frac{d}{dz} + (z - \alpha) \frac{\dot{g}^{(\alpha)}_m}{g^{(\alpha)}_m} + 
  z \frac{\ddot{g}^{(\alpha)}_m}{g^{(\alpha)}_m}\right] y^{(\alpha)}_n(z) = (m - n)
  y^{(\alpha)}_n(z). \label{eq:eq-y} 
\end{equation}
\par
%
%
The orthonormality and completeness of $\psi^{(-)}_{\nu}(x)$, $\nu=0$, 1, 2,~\ldots, on the half-line imply that the polynomials $y^{(\alpha)}_n(z)$, $n = m+\nu$, $\nu=0$, 1, 2,~\ldots,  form an orthogonal and complete set  with respect to the positive-definite measure $z^{\alpha} e^{-z} \bigl(g^{(\alpha)}_m\bigr)^{-2} dz$. According to the choice made for $\phi$, such polynomials belong to the $L1$ or $L2$ family of $X_m$ EOP and are denoted by $L^{\rm I}_{\alpha, m, n}(z)$ or $L^{\rm II}_{\alpha, m, n}(z)$. In \cite{odake09a, odake10a, ho09}, they were normalized in such a way that their highest-degree term is given by $(-z)^n/[(n-m)! m!]$ multiplied by $(-1)^m$ or 1, respectively. In conformity with the choices made in \cite{gomez10a, gomez09, cq11b}, we will drop the additional factor $(-1)^m$ in case I. Then $L^{\rm I}_{\alpha,1,n}(z)$ and $L^{\rm II}_{\alpha,1,n}(z)$ both exactly reduce to $\hat{L}^{(\alpha)}_n(z)$, considered in equations (\ref{eq:ex-X1}), (\ref{eq:eq-X1}) and (\ref{eq:int-X1}). With this convention, it is easy to see that
\begin{equation}
  L^{\rm I}_{\alpha,m,n}(z) = (-1)^{m-1} \left[g^{(\alpha)}_m \left(\frac{d}{dz} - 1\right) - \dot{g}^{(\alpha)}_m
  \right] L^{(\alpha-1)}_{\nu}(z),  \label{eq:LI-L}
\end{equation}
\begin{equation}
  L^{\rm II}_{\alpha,m,n}(z) = (\alpha + \nu + 1 - m)^{-1} \left[g^{(\alpha)}_m \left(z\frac{d}{dz} + \alpha + 1
  \right) - z\dot{g}^{(\alpha)}_m\right] L^{(\alpha+1)}_{\nu}(z),  \label{eq:LII-L}
\end{equation}
by comparing the highest-degree terms on both sides of the equations.\par
%
%
In the present approach, to get $V_{l,{\rm ext}}(x)$ we have had to reparametrize the conventional potential by considering $V_{l'}(x)$ with $l' \ne l$. It is worth mentioning that an isospectral transformation from $V_l(x)$ to $V_{l,{\rm ext}}(x)$ with the same parameter can be obtained in the framework of second-order SUSYQM \cite{bagchi09a, cq09}.\par
%
%
\section{\boldmath Rationally-extended radial oscillator and Laguerre-type $X_{m_1,m_2}$ EOP in second-order SUSYQM}

To illustrate the construction of multi-indexed families of EOP and corresponding rationally-extended potentials, let us consider the case of the radial oscillator potential in second-order SUSYQM (SSUSY) \cite{cq11a}.\par
%
%
In such a setting, one starts from a pair of Hamiltonians \cite{andrianov, bagrov, samsonov, bagchi99, aoyama, fernandez}
\begin{equation}
  H^{(1)} = - \frac{d^2}{dx^2} + V^{(1)}(x), \qquad H^{(2)} = - \frac{d^2}{dx^2} + V^{(2)}(x), 
\end{equation}
intertwining with two second-order differential operators
\begin{equation}
  {\cal A}^{\dagger} = \frac{d^2}{dx^2} - 2p(x) \frac{d}{dx} + q(x), \qquad {\cal A} = \frac{d^2}{dx^2} + 2p(x)
  \frac{d}{dx} + 2p'(x) + q(x),
\end{equation}
as ${\cal A} H^{(1)} = H^{(2)} {\cal A}$ and ${\cal A}^{\dagger} H^{(2)} = H^{(1)} {\cal A}^{\dagger}$. These intertwining relations imply that the functions $p(x)$, $q(x)$, $V^{(1)}(x)$ and $V^{(2)}(x)$ are constrained by the relations
\begin{equation}
\begin{split}
  & q(x) = - p' + p^2 - \frac{p''}{2p} + \left(\frac{p'}{2p}\right)^2 - \frac{c^2}{16p^2},  \\
  & V^{(1,2)}(x) = \mp 2p' + p^2 + \frac{p''}{2p} - \left(\frac{p'}{2p}\right)^2 + \frac{c^2}{16p^2}, 
\end{split}  
\end{equation}
where $c$ is some integration constant. Hence the knowledge of $p(x)$ and $c$ determines everything else. SSUSY offers more possibilities of manipulating the initial spectrum than standard SUSYQM, such as deleting or creating one or two levels, moving one level or leaving the spectrum unchanged. Here we shall only be concerned with the last eventuality (isospectral transformation).\par
%
%
In the reducible case to be considered here, ${\cal A}^{\dagger}$ and $\cal A$ can be factorized into a product of two first-order differential operators, ${\cal A}^{\dagger} = A^{\dagger} \tilde{A}^{\dagger}$ and ${\cal A} = \tilde{A} A$, where $A^{\dagger}$ and $A$ are expressed as in (\ref{eq:SUSYQM}), while
\begin{equation}
  \tilde{A}^{\dagger} = - \frac{d}{dx} + \tilde{W}(x), \qquad \tilde{A} = \frac{d}{dx} + \tilde{W}(x).
\end{equation}
Then $A^{\dagger}$ and $A$ may be used as ladder operators in first-order SUSYQM with $H^{(+)} = A^{\dagger} A$ and $H^{(-)} = A A^{\dagger}$ as in (\ref{eq:SUSYQM}), while $W(x) = - \bigl(\log \phi(x)\bigr)'$ and $H^{(+)} \phi(x) = 0$. Similarly, from $\tilde{A}^{\dagger}$ and $\tilde{A}$ one may build $\tilde{H}^{(+)} = \tilde{A}^{\dagger} \tilde{A}$ and $\tilde{H}^{(-)} = \tilde{A} \tilde{A}^{\dagger}$, corresponding to some potentials $\tilde{V}^{(\pm)}(x)$, factorization function $\tilde{\phi}(x)$ and factorization energy $\tilde{E}$, such that $\tilde{V}^{(\pm)}(x) = \tilde{W}^2(x) \mp \tilde{W}'(x) + \tilde{E}$, $\tilde{W}(x) = - \bigl(\log \tilde{\phi}(x)\bigr)'$ and $\tilde{H}^{(+)} \tilde{\phi}(x) = 0$. Whenever $\tilde{V}^{(+)}(x) = V^{(-)}(x)$, so that both pairs $\bigl(H^{(+)}, H^{(-)}\bigr)$ and $\bigl(\tilde{H}^{(+)}, \tilde{H}^{(-)}\bigr)$ can be glued together, we get a reducible SSUSY system or, equivalently, a second-order parasupersymmetric (PSUSY) one \cite{rubakov, khare93a}.\par
%
%
In the isospectral case, we have to assume $E < E^{(+)}_0$, $\tilde{E} < \tilde{E}^{(+)}_0 = E^{(+)}_0$ and to demand that both $\phi^{-1}(x)$ and $\tilde{\phi}^{-1}(x)$ be nonnormalizable (together with $\phi(x)$ and $\tilde{\phi}(x)$).\par
%
%
The connection between the SSUSY and PSUSY approaches is given by the relations
\begin{equation}
\begin{split}
  & H^{(1)} = H^{(+)} + \tfrac{c}{2}, \qquad H^{(2)} = \tilde{H}^{(-)} - \tfrac{c}{2}, \qquad H = H^{(-)} 
       + \tfrac{c}{2}= \tilde{H}^{(+)} - \tfrac{c}{2}, \\
  & p(x) = \tfrac{1}{2} \bigl(W + \tilde{W}\bigr), \qquad c = E - \tilde{E},
\end{split}
\end{equation}
showing that the intermediate Hamiltonian $H$ is both partner to $H^{(1)}$ and $H^{(2)}$.\par
%
%
Instead of $\phi(x)$ and $\tilde{\phi}(x)$, we may start from two eigenfunctions $\phi_1(x)$ and $\phi_2(x)$ of the same $V^{(+)}(x)$ with respective energies $E_1$ and $E_2$ (less than $E^{(+)}_0$) and such that $\phi_1^{-1}$ and $\phi_2^{-1}$ are nonnormalizable. Then, $\phi = \phi_1$, $\tilde{\phi} = A \phi_2 = {\cal W}(\phi_1, \phi_2)/\phi_1$, $E = E_1$ and $\tilde{E} = E_2$, where ${\cal W}(\phi_1, \phi_2)$ denotes the Wronskian of $\phi_1(x)$ and $\phi_2(x)$. As a consequence,
\begin{equation}
  p(x) = - \frac{{\cal W}'(\phi_1, \phi_2)}{2 {\cal W}(\phi_1, \phi_2)} = - \frac{(E_1 - E_2) \phi_1 \phi_2}
 {2 {\cal W}(\phi_1, \phi_2)}, 
\end{equation}
which entirely determines the SSUSY partner potential through the equation
\begin{equation}
  V^{(2)}(x) = V^{(1)}(x) + 4p'(x).
\end{equation}  
\par
%
%
In the case of the radial oscillator potential, we know that there are two types of seed functions (\ref{eq:phi-I}) and (\ref{eq:phi-II}), which may be used. As the order of $\phi_1$ and $\phi_2$ is irrelevant as far as the final potential $V^{(2)}(x)$ is concerned, there are three types of possibilities for the pair $(\phi_1, \phi_2)$. In all three cases, we have to start from a potential $V^{(+)}(x) = V_{l'}(x)$ with some different $l'$. The Wronskian ${\cal W}(\phi_1(x), \phi_2(x))$ can be written in terms of some $\mu$th-degree polynomial in $z$, $g^{(\alpha)}_{\mu}(z)$, itself expressible in terms of a Wronskian $\tilde{{\cal W}}(f(z), g(z))$ of two appropriate functions of $z$. The results read
\begin{eqnarray}
  & (i) \; &  V^{(+)} = V_{l-2}, \quad \phi_1 = \phi^{\rm I}_{l-2,m_1}, \quad \phi_2 = \phi^{\rm I}_{l-2,m_2}, 
         \quad 0 < m_1 < m_2, \nonumber \\
  & & {\cal W}(\phi_1, \phi_2) = \omega x (\chi^{\rm I}_{l-2})^2 g^{(\alpha)}_{\mu}(z), \nonumber \\
  & & g^{(\alpha)}_{\mu}(z) = \tilde{{\cal W}}\bigl(L^{(\alpha-2)}_{m_1}(-z), L^{(\alpha-2)}_{m_2}(-z)\bigr),
         \quad \mu = m_1 +  m_2 - 1; \label{eq:i}  \\
  & (ii) \; & V^{(+)} = V_{l+2}, \quad \phi_1 = \phi^{\rm II}_{l+2,m_1}, \quad \phi_2 = \phi^{\rm II}_{l+2,m_2}, 
         \quad 0 < m_1 < m_2 < \alpha + 2, \nonumber \\
  & & {\cal W}(\phi_1, \phi_2) = \omega x (\chi^{\rm II}_{l+2})^2 g^{(\alpha)}_{\mu}(z), \nonumber \\
  & & g^{(\alpha)}_{\mu}(z) = \tilde{{\cal W}}\bigl(L^{(-\alpha-2)}_{m_1}(z), L^{(-\alpha-2)}_{m_2}(z)\bigr), 
         \quad \mu = m_1 + m_2 - 1; \label{eq:ii}  \\ 
  &  (iii) \; & V^{(+)} = V_l, \quad \phi_1 = \phi^{\rm I}_{l,m_1}, \quad \phi_2 = \phi^{\rm II}_{l,m_2}, 
        \quad 0 < m_1, \quad 0 < m_2 < \alpha, \nonumber \\
  & & {\cal W}(\phi_1, \phi_2) = \frac{2}{x} \chi^{\rm I}_l \chi^{\rm II}_l g^{(\alpha)}_{\mu}(z), \nonumber \\
  & & g^{(\alpha)}_{\mu}(z) = z \tilde{{\cal W}}\bigl(L^{(\alpha)}_{m_1}(-z), L^{(-\alpha)}_{m_2}(z)\bigr) 
        - (z + \alpha) L^{(\alpha)}_{m_1}(-z) L^{(-\alpha)}_{m_2}(z), \nonumber \\
  & & \mu = m_1 + m_2 + 1.  \label{eq:iii}
\end{eqnarray}
\par
%
%
In all three cases, provided $g^{(\alpha)}_{\mu}(z)$ does not vanish on the half-line, the SSUSY partner potentials can be written as
\begin{equation}
\begin{split}
  V^{(1)} & = V_{l'} - \frac{1}{2} (E_1 + E_2), \\
  V^{(2)} & = V_l - 2\omega \left\{\frac{\dot{g}^{(\alpha)}_{\mu}}{g^{(\alpha)}_{\mu}} + 2z 
        \left[\frac{\ddot{g}^{(\alpha)}_{\mu}}{g^{(\alpha)}_{\mu}}
        - \left(\frac{\dot{g}^{(\alpha)}_{\mu}}{g^{(\alpha)}_{\mu}}\right)^2\right]\right\} 
        - \frac{1}{2} (E_1 + E_2) + C, 
\end{split}  \label{eq:SSUSY-pot}
\end{equation}
where $C = - 2\omega$, $2\omega$ or 0 in case (i), (ii) or (iii), respectively. Hence, up to some additive constant, they assume essentially the same form as $V^{(+)}$ and $V^{(-)}$ in first-order SUSYQM (see equation (\ref{eq:V-rat})) with the only exception that $g^{(\alpha)}_m(z)$ is now replaced by $g^{(\alpha)}_{\mu}(z)$.\par
%
%
In the PSUSY setting, the potential in the intermediate $H$ is some $V_{l-1, {\rm ext}}$, $V_{l+1, {\rm ext}}$ or $V_{l+1, {\rm ext}}$, as obtained in first-order SUSYQM. If we change the order of $\phi_1$ and $\phi_2$, then the intermediate potential is different although the final one remains the same. In particular, in case (iii), it becomes a $V_{l-1, {\rm ext}}$ potential.\par
%
%
It is worth noting that extending the range of $m_1$, $m_2$ values to include $m_1=0$ in case (i) or (ii) and $m_1=0$ or $m_2=0$ in case (iii) would not lead to any new result, because the corresponding $g^{(\alpha)}_{\mu}(z)$ would then reduce to some $g^{(\alpha)}_m(z)$ already found in first order. Even for the $m_1$ and $m_2$ values listed in (\ref{eq:i}), (\ref{eq:ii}) and (\ref{eq:iii}), there are some possibilities of coincidence. As a matter of fact, the lowest-degree new $g^{(\alpha)}_{\mu}(z)$ polynomial corresponds to $\mu=3$ and is given by $g^{(\alpha)}_3(z) = [z^3 + 3\alpha z^2 + 3(\alpha-1)(\alpha+1) z + (\alpha-1)\alpha(\alpha+1)]/3$. It may be obtained in case (i) or (ii) for $m_1=1$, $m_2=2$, or else in case (iii) for $m_1=m_2=1$. Up to second order, there are therefore three distinct cubic-type extended potentials, the first two being those found in first order and associated with either $L^{(\alpha-1)}_3(-z)$ or $L^{(-\alpha-1)}_3(z)$, the third one being that corresponding to the above-mentioned $g^{(\alpha)}_3(z)$. Since this conclusion remains true when considering $k$th-order SUSYQM with $k>2$, it has been conjectured that there are exactly $\mu$ distinct extended radial oscillator potentials (and corresponding EOP families) of $\mu$th type \cite{cq11a}. This assertion has actually been proved for $\mu=1$, 2 and 3.\par
%
%
Turning now ourselves to the wavefunctions $\psi^{(2)}_{\nu}(x)$ of $H^{(2)}$, we note that we may obtain them from those of $H^{(1)}$, $\psi^{(1)}_{\nu}(x) \propto \eta_{l'}(z) L^{(\alpha')}_{\nu}(z)$, $\nu=0$, 1, 2, \ldots, by applying the second-order differential operator $\cal A$. They can be written as
\begin{equation}
  \psi^{(2)}_{\nu}(x) \propto \frac{\eta_l(z)}{g^{(\alpha)}_{\mu}(z)} y^{(\alpha)}_n(z), \qquad n = \mu + \nu,
  \qquad \nu=0, 1, 2, \ldots,
\end{equation}
where $y^{(\alpha)}_n(z)$ is some $n$th-degree polynomial in $z$, satisfying a second-order differential equation similar to equation (\ref{eq:eq-y}), but with $\mu$ and $g^{(\alpha)}_{\mu}(z)$ substituted for $m$ and $g^{(\alpha)}_m(z)$, respectively. As the polynomials considered in section 3, those obtained here form orthogonal and complete sets  with respect to a positive-definite measure of the type $z^{\alpha} e^{-z} \bigl(g^{(\alpha)}_{\mu}\bigr)^{-2} dz$. We have therefore got (in general) $X_{m_1,m_2}$ EOP belonging to three families, which may be denoted by $L^{\rm I, I}_{\alpha, m_1, m_2, n}(z)$, $L^{\rm II, II}_{\alpha, m_1, m_2, n}(z)$ and $L^{\rm I, II}_{\alpha, m_1, m_2, n}(z)$, respectively. We choose to normalize them in such a way that their highest-degree term is equal to $(-z)^n/[(n-\mu)!m_1!m_2!]$.\par
%
%
{}From the factorized form ${\cal A} = \tilde{A} A$, it can be shown that the $X_{m_1,m_2}$ EOP can be obtained from the $X_{m_1}$ EOP by applying some first-order differential operator, thereby generalizing equations (\ref{eq:LI-L}) and (\ref{eq:LII-L}) \cite{cq11b}. The results read
\begin{equation}
  L^{\rm I,I}_{\alpha, m_1,m_2,n}(z) = (-1)^{m_2} (m_2-m_1)^{-1} {\cal O}^{(\alpha)}_{m_1,\mu}
  L^{\rm I}_{\alpha,m_1,n_1}(z),  \label{eq:LI-I}
\end{equation}
\begin{equation}
  L^{\rm II,II}_{\alpha, m_1,m_2,n}(z) = - [(\alpha+\nu+2-m_2) (m_2-m_1)]^{-1} {\cal O}^{(\alpha)}_{m_1,\mu}
  L^{\rm II}_{\alpha,m_1,n_1}(z),  \label{eq:LII-II}
\end{equation}
\begin{equation}
  L^{\rm I,II}_{\alpha, m_1,m_2,n}(z) = (\alpha+\nu-m_2)^{-1} {\cal O}^{(\alpha)}_{m_1,\mu}
  L^{\rm I}_{\alpha,m_1,n_1}(z),  \label{eq:LI-II}
\end{equation}
where $n=\mu+\nu$, $n_1=m_1+\nu$, $\nu=0$, 1, 2,~\ldots, and
\begin{equation}
  {\cal O}^{(\alpha)}_{m_1,\mu} = \frac{1}{g^{(\alpha-1)}_{m_1}} \left[g^{(\alpha)}_{\mu} \left(\frac{d}{dz}
  - 1\right) - \dot{g}^{(\alpha)}_{\mu}\right]
\end{equation}
in (\ref{eq:LI-I}), while
\begin{equation}
  {\cal O}^{(\alpha)}_{m_1,\mu} = \frac{1}{g^{(\alpha+1)}_{m_1}} \left[g^{(\alpha)}_{\mu} \left(z\frac{d}{dz}
  + \alpha + 1\right) - z\dot{g}^{(\alpha)}_{\mu}\right]
\end{equation}
in (\ref{eq:LII-II}) or (\ref{eq:LI-II}).\par
%
%
The approach used in this section has been extended to $k$th-order SUSYQM \cite{cq11b}, in which case one may distinguish $k+1$ families of $X_{m_1,m_2,\ldots,m_k}$ EOP, denoted by $L^{(q,k-q)}_{\alpha,m_1,m_2,\ldots,m_k,n}(z)$ and associated with $k+1$ possible choices I$^q$II$^{k-q}$, $0 \le q \le k$, for the set of $k$ seed functions of type I or II. Here it is assumed that $0 < m_1 < m_2 < \cdots < m_q$ correspond to type I, whereas $0 < m_{q+1} < m_{q+2} < \cdots < m_k$ belong to type II. The EOP degree $n$ has been shown to be given by $n=\mu+\nu$, $\nu=0$, 1, 2,~\ldots, with
\begin{equation}
  \mu = \sum_{i=1}^k m_i - \frac{1}{2} q(q-1) - \frac{1}{2} (k-q)(k-q-1) + q(k-q).
\end{equation}
\par
%
%
\section{Shape invariance of rationally-extended potentials}

In first-order SUSYQM, if one knows the spectrum and wavefunctions of $V^{(+)}(x)$ (resp.\ $V^{(-)}(x)$), then one is able to obtain those of its partner $V^{(-)}(x)$ (resp.\ $V^{(+)}(x)$) by application of $A$ (resp.\ $A^{\dagger}$). For some special class of potentials, however, no previous knowledge is needed to be able to derive the spectrum and wavefunctions of both partners. This is the case of so-called SI potentials \cite{gendenshtein}, i.e., those potentials for which the two partners are similar in shape and differ only in the parameters. More precisely, $V^{(\pm)}(x;a_1)$ are SI if
\begin{equation}
  V^{(-)}(x;a_1) = V^{(+)}(x;a_2) + R(a_1),  \label{eq:SI}
\end{equation}
where $a_1$ is a set of parameters, $a_2$ is a function of $a_1$ (i.e., $a_2 = f(a_1)$) and the remainder $R(a_1)$ is independent of $x$ \cite{cooper00}. When combined with unbroken SUSYQM (case i or iii below equation (\ref{eq:phi-E})), equation (\ref{eq:SI}) indeed provides an integrability condition so that the two potentials are solvable by purely algebraic methods. Such a condition is only a sufficient one, since many ES potentials that are not SI are known.\par
%
%
In case i, for instance, equation (\ref{eq:SI}) can be rewritten as
\begin{equation}
  W^2(x;a_1) + W'(x;a_1) = W^2(x:a_2) - W'(x;a_2) + R(a_1),  \label{eq:SI-bis}
\end{equation}
where $W(x;a_1) = - \bigl(\log \psi^{(+)}_0(x;a_1)\bigr)'$ in terms of the ground-state wavefunction $\psi^{(+)}_0(x;a_1)$ of $V^{(+)}(x;a_1)$. Since equation (\ref{eq:SI-bis}) is a difference-differential equation relating the square of the superpotential $W$ and its spatial derivative computed at two sets $a_1$, $a_2$ of parameter values, the classification of its solutions is not an easy task.\par
%
%
The simplest case corresponds to translationally SI potentials for which the parameters $a_1$ and $a_2$ are related to each other by translation ($a_2 = a_1 + \lambda$) \cite{carinena00, cooper87, dabrowska}. Other cases have also been studied, such as SI potentials under scaling \cite{khare93b, barclay} and cyclic SI potentials \cite{sukhatme}. In the last cases, the potentials are only known formally as Taylor series, whereas in the first one, they are explicitly known in a closed form in terms of simple functions.\par
%
%
{}For the conventional radial oscillator $V_l(x)$, whose ground-state wavefunction is given by $\psi^{(l)}_0(x) \propto \eta_l(z)$ with $\eta_l(z)$ defined in (\ref{eq:z}), the role of $a_1$ is played by $l$ and the superpotential $W(x;l) = W_l(x)$ can be written as 
\begin{equation}
  W_l(x) = \frac{1}{2} \omega x - \frac{l+1}{x} = - \omega x \frac{\dot{\eta}_l}{\eta_l}.
\end{equation}
As expected, $V_l(x) - E^{(l)}_0 = W_l^2(x) - W'_l(x)$, where $E^{(l)}_0 = \omega \bigl(l + \frac{3}{2}\bigr) = \omega (\alpha+1)$. In such a case, equation (\ref{eq:SI-bis}) translates into
\begin{equation}
  W_l^2(x) + W'_l(x) = W_{l+1}^2(x) - W'_{l+1}(x) + 2\omega = \Bigl[V_{l+1}(x) - E^{(l+1)}_0\Bigr] +
  2\omega,  \label{eq:SI-osc}
\end{equation}
which illustrates the well-known SI of $V_l(x)$, its partner being $V_{l+1}(x)$ (hence $a_2 = l+1$) and the remainder $2\omega$ being independent of $x$ \cite{cooper00}.\par
%
%
It turns out that the corresponding rationally-extended potentials $V_{l,{\rm ext}}(x) = V_l(x) + V_{l,{\rm rat}}(x)$, where $V_{l,{\rm rat}}(x)$ is expressed in terms of some (nonvanishing) $\mu$th-degree polynomial $g^{(\alpha)}_{\mu}(z)$ as in equation (\ref{eq:SSUSY-pot}), have very similar properties \cite{cq08, cq09, odake09a, odake10a, odake10b, ho09, gomez10b, grandati11a, odake11a, cq11b, grandati11b}. This is actually a consequence of the fact that their ground-state wavefunction can be written as $\bar{\psi}^{(l)}_0(x) \propto \eta_l(z) g^{(\alpha+1)}_{\mu}(z) / g^{(\alpha)}_{\mu}(z)$ or, in other words, that the lowest-degree EOP associated with $g^{(\alpha)}_{\mu}(z)$ is $y^{(\alpha)}_{\mu}(z) \propto g^{(\alpha+1)}_{\mu}(z)$. From $\bar{\psi}^{(l)}_0(x)$, we indeed get the superpotential
\begin{equation}
\begin{split}
  & \bar{W}_l(x) = \bar{W}_{l,1}(x) + \bar{W}_{l,2}(x), \\
  & \bar{W}_{l,1}(x) = W_l(x), \qquad \bar{W}_{l,2}(x) = - \omega x 
     \left(\frac{\dot{g}^{(\alpha+1)}_{\mu}}{g^{(\alpha+1)}_{\mu}} - \frac{\dot{g}^{(\alpha)}_{\mu}}
     {g^{(\alpha)}_{\mu}}\right),
\end{split}  \label{eq:W-ext}
\end{equation}
which satisfies the equation $V_{l,{\rm ext}}(x) - E^{(l)}_0 = \bar{W}_l^2(x) - \bar{W}'_l(x)$, as it should be. Then, with $a_1 = l$, $a_2 = l+1$ and $R(a_1) = 2\omega$, equation (\ref{eq:SI-bis}) becomes
\begin{equation}
  \bar{W}_l^2(x) + \bar{W}'_l(x) = \bar{W}_{l+1}^2(x) - \bar{W}'_{l+1}(x) + 2\omega = 
  \Bigl[V_{l+1,{\rm ext}}(x) - E^{(l+1)}_0\Bigr] + 2\omega,
\end{equation}
which is a counterpart of equation (\ref{eq:SI-osc}).\par
%
%
Since similar results have been obtained for extended potentials associated with Jacobi EOP, we conclude that the extensions considered here do not spoil the nice SI property of the conventional potentials, from which they are built. A whole new class of so far unknown translationally SI potentials has therefore been created.\par
%
%
The appearance of such a novel family has motivated two recent re-examinations of the SI condition (\ref{eq:SI-bis}), independently of the EOP construction problem, in the case where the parameters are linked by translation.\par
%
%
In the first one \cite{bougie10, bougie11}, the non-local differential equation for $W$ has been replaced by two local partial differential equations. The first of them is similar to the Euler equation expressing momentum conservation for inviscid fluid flow in one spatial dimension, while the second provides a constraint helping one to determine unique solutions. In units wherein $\hbar$ is not set equal to one (in contrast with what is done here and in most other works), all the solutions of these equations, which do not depend explicitly on $\hbar$, have been derived and proved to include all conventional translationally SI potentials and nothing else. The generalization of the algorithm to superpotentials that contain $\hbar$ explicitly has given rise to some of the extended potentials previously built in connection with EOP \cite{cq08, bagchi09a}.\par
%
%
In the second study \cite{ramos}, as in equation (\ref{eq:W-ext}), the superpotential is separated into two parts, the first one being the superpotential associated with some conventional potential. The second part is then divided into two pieces again and the latter are shown to satisfy some compatibility condition, for which a few solutions are provided. These give rise to already known extended potentials \cite{cq08, bagchi09a} or slight generalizations thereof.\par
%
%
Although the two last works are interesting because they have shed some light on the resolution of the SI condition, they have not led up to now to any really novel example of translationally SI rationally-extended potential (although the authors of \cite{bougie11, ramos} claim the opposite). From this viewpoint, the technique based on EOP construction seems to be more powerful. It has the additional advantage of simultaneously providing the solutions of the Schr\"odinger equation at a low price.\par
%
%
\section{Final comments}

The purpose of this concluding section is to briefly review some works connected somehow or other with the topics of EOP and of related ES quantum potentials.\par
%
%
Some applications of EOP have been considered to other problems than the resolution of the conventional Schr\"odinger equation. The inclusion of a position-dependent effective mass in the latter has been dealt with \cite{midya}, as well as the replacement of the real potential by a complex one, either $\cal PT$-symmetric or not \cite{bagchi09a, bagchi10a}. The EOP have been studied in the framework of the quantum Hamilton-Jacobi formalism in connection with the supersymmetric WKB quantization condition \cite{sree} and in that of type A $\cal N$-fold supersymmetry and quasi-solvability \cite{bagchi09b, tanaka, bagchi10b}. They have also been shown to be useful in the context of the Dirac equation coupled minimally or non-minimally with some external field and in that of the Fokker-Planck equation \cite{ho11b}.\par
%
%
Some generalizations of the $X_m$ EOP for noninteger $m$, which are not polynomials but are expressible in terms of the (confluent) hypergeometric function, have been proved to lead to SI potentials \cite{odake11b}. Some of the latter had been studied before and shown to belong to the class of conditionally ES potentials, although their SI had not been recognized \cite{junker97, junker98, dutta}.\par
%
%
In the course of the construction of the first examples of type I and type II Laguerre and Jacobi $X_2$ EOP \cite{cq09}, there occurred so-called type III polynomials, which were generalized later on to higher $m$ values, hence giving rise to what is sometimes referred to as the L3 and J3 families \cite{grandati11a, ho11a, garcia, grandati11c, ho11c}. These polynomials, however, do not qualify to be called EOP, since some degrees are missing from the set of them, and the corresponding extended potentials are not SI either. They belong instead to classes of polynomials and potentials often encountered (and periodically rediscovered) in applications (see, e.g., \cite{gomez04a, junker97, carinena08, fellows}).\par
%
%
{}Finally, it is worth mentioning  that families of EOP have also been constructed in connection with Wilson, Askey-Wilson, Racah and $q$-Racah polynomials used in the framework of discrete quantum mechanics \cite{garcia, odake09b, odake10c, odake11c}.\par
%
%
\section*{Acknowledgments}

The author would like to thank Y Grandati for several useful discussions.\par
%
%

\end{document}